\documentclass{article}
 \usepackage[dvips]{graphics}

\newtheorem{thrm}{Proposition}

\newcommand{\be}{\begin{equation}}
\newcommand{\bea}{\begin{eqnarray}}
\newcommand{\ba}{\begin{array}}
\newcommand{\ee}{\end{equation}}
\newcommand{\eea}{\end{eqnarray}}
\newcommand{\ea}{\end{array}}
\newcommand{\bean}{\begin{eqnarray*}}
\newcommand{\eean}{\end{eqnarray*}}
\newcommand{\rmd}{{\rm d}}
\newcommand{\openone}{1}
\newcommand{\abs}[1]{\left| #1 \right|}
\newcommand{\bc}{\begin{center}}
\newcommand{\ec}{\end{center}}

\begin{document}

\title{Are Simple Real Pole Solutions Physical?}

\author{M. Berg\footnote{This work was
    partially performed while
    visiting the Dept.\ of Plasma Physics
    at Ume{\aa} University.} \footnote
    {Electronic mail: mberg@einstein.ph.utexas.edu} \\
    {\it Center for Relativity} \ \\  {\it University
    of Texas at Austin, USA} 
   \\[3mm]
   M. Bradley\footnote{Electronic mail:
   michael.bradley@physics.umu.se} \\
   {\it Department of Plasma Physics}  \\
   {\it Ume{\aa} University, Ume{\aa}, Sweden} \\ \ \\
}

\maketitle

\centerline{PACS Ref: 04.20.Jb, 04.30.Nk, 04.40.Nr}

\begin{abstract}
We consider exact solutions generated by
the inverse scattering technique,
also known as the soliton transformation.
In particular, we study the class of simple real pole solutions.
For quite some time, those solutions have been considered interesting as
models of cosmological shock waves.
A coordinate singularity on the wave fronts
was removed by a transformation which induces a 
null fluid with negative energy density on the wave front.
This null fluid is usually seen as another coordinate artifact,
since there seems to be a general belief that that this kind of solution 
can be seen as the real pole limit of the
smooth solution generated with a pair of complex conjugate poles in the
transformation. We perform this limit explicitly,
and find that the belief is unfounded:
two coalescing complex conjugate poles cannot yield
a solution with one real pole.
Instead, the two complex conjugate poles 
go to a different limit, what we call a ``pole on a pole''. 
The limiting procedure is not unique; it is
sensitive to how quickly some parameters approach zero.
We also show that there exists
no improved coordinate transformation which would
remove the negative energy density. We conclude that
negative energy is an intrinsic part of this class of solutions.
\end{abstract}

\hfill gr-qc/0002042\\
              
\section{Introduction}
\subsection{Background}
We consider exact solutions generated by
the inverse scattering technique,
also known as the soliton transformation, in the
so-called cosmological case (see section \ref{sec:setup} below).
We do not consider the stationary axisymmetric case 
since we are interested in solutions of wave-like
character and cosmological application.
A comprehensive review of this subject and applications
related to that of this paper has been given by Verdaguer
\cite{Verd}. In the review one also finds references to
earlier work in the field by Ernst, Chandrasekhar and others
in the 1970s and 1980s.
A general discussion of
waves modelled by exact solutions
is given in the book by Griffiths \cite{Griffiths}.

The inverse scattering technique was first applied to 
general relativity in 1978 by Belinskii and
Zakharov \cite{Beli2}.
Their algorithm generates exact vacuum solutions
to Einstein's equations from a known ``seed solution'',
also called the ``background''.
This background is multiplied by the so-called {\it scattering
matrix} to yield the new metric, called the perturbation.
Note that the use of the word perturbation does not 
imply an approximation,
the new metric is still an exact
solution.
Any poles in the scattering
matrix give rise to solitonic solutions,
and these have earned the inverse scattering
technique its other name ``soliton transformation''.
The transformation is closely related to the
{\it B\"{a}cklund transformations}, and the exact
relation has been clarified by Cosgrove\cite{Cosgrove1,Cosgrove2}.

\subsection{Simple real pole solutions}

The inverse scattering technique can recreate 
well-known solutions. One interesting example is
the Kerr-NUT metric generated from flat space
\cite{Beli3} with two
real poles in the scattering matrix.
Homogenous cosmologies of Bianchi types I through VII
can be generated with the inverse scattering technique.
New solutions have also been produced, often with 
gravitational solitons.
One class of such solutions that has been studied extensively 
\cite{Verd,Beli2,Franc,Brad1,FrancCurir} is the ``one real
pole'' perturbation on a Bianchi I or II background
(for the purposes of this article, we choose the Kasner vacuum
metric as a background). One real pole solutions
may have merits as cosmological models; to see this we 
briefly survey the physical behavior of the solutions.

\begin{figure}[h]
   \begin{center}
     \resizebox{5cm}{!}{\includegraphics{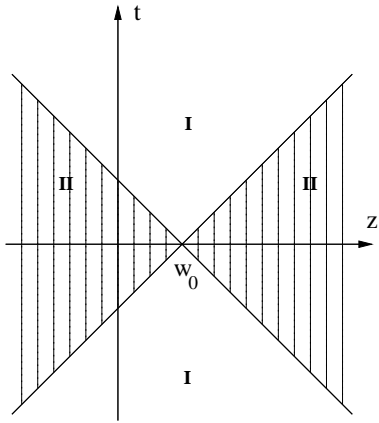}}
     \caption{The light cone structure. Pole at $w_0$.}
     \label{fig:cone}
  \end{center}
\end{figure}

The relevant region for cosmology is $t>0$ (see fig. \ref{fig:cone}).
A solitary shock wave emerges from the initial singularity
at $z=w_0$, $t=0$.
Behind the wavefront, the shock wave leaves
way for the background (Kasner) metric. 

Thus this soliton-like disturbance
sweeps away some initial disorder which
surrounds the singularity. This has earned the single
real pole solution generated from a Bianchi I or II metric its nickname
``the cosmic broom''.
(From the Italian {\it la scopa cosmica}.)
Simple real pole solutions generated from other seed metrics
or with more poles display similar behavior. 
The cosmic broom represents an interesting exact-solution model of
gravitational waves of cosmological origin. A background of such waves
could in principle be detected, and the cosmic broom would provide
an intriguing classical
alternative to waves generated from the initial conditions of the
universe in quantum cosmology.

Furthermore, it turns out that the perturbation is
undefined in region I,
so we match the perturbation in region II 
across the light
cone to the background (Kasner) metric in region I.
While other choices for the matching are possible,
we match to the background for two main reasons. One reason 
is that the perturbed
metric in region II itself goes into the unperturbed
one when approaching the light cone.
The other reason is that we wish to
maintain the traveling wave interpretation, which
is most meaningful when the metric is asymptotically 
equal to the background on {\em both} sides of the disturbance.

In general, one real pole is not sufficient to recover the background at
spatial infinity. Instead, this can be achieved with two or more distinct
real poles \cite{Verd}.
Since each real pole defines a light cone, one then has to
perform a number of matchings. Some of these matchings will involve
matching the background within the inner light-cone to one-pole solutions.
Consequently, a first step is to match the background to a one-pole
solution, even though one is ultimately interested in $n$-pole solutions
or a stochastic background of waves. Therefore, the problem described in
this paper can be seen as part of the more complicated matching
procedure for several distinct real poles. 

A good motivation to use a Kasner-type background
is that it has been shown that any ``generic''
singularity is followed by a succession of Kasner eras \cite{Beli1}.
Moreover, a Bianchi I metric is approximately unaffected
close to the singularity by the presence of matter
\cite{Stephani}.
(Recently,
doubt has been cast on this somewhat folkloristic 
claim \cite{Uggla}.)

\subsection{Problems}
The cosmic broom suffers from metric coefficient
infinities on the light cone in canonical 
(Belinskii-Zakharov) coordinates.
Recently it was shown that these coordinate infinities
can be removed by a certain (singular) coordinate transformation
\cite{Brad1,FrancCurir}. However,
a new but potentially less serious problem was created by
this transformation. Although the {\it metric} is now continuous,
the Ricci tensor goes distributional-valued, 
i.e. acquires $\delta$-functions, at the light cone.
This kind of $\delta$-function can normally be interpreted
as an impulsive wave, e.g. a null fluid.
(see the book by Griffiths\cite{Griffiths} on the subject).
Unfortunately, the null fluid in this case
would have {\it negative} energy density \cite{Brad1}.
Let us exhibit this explicitly for the
left hand side of the light cone ($z<w_0$), a similar
discussion holds for the right hand side.
{}From Einstein's equation with the aforementioned
Ricci tensor one finds
\[
T_{ij}=-{1 \over 8\pi t \, g_{tt} }\, \delta\, (t+z-w_0) \; v_i \, v_j
\]
where $v^a = 1/\sqrt{2g_{tt}} \, (\partial/\partial t-
\partial/\partial z)$ is along the light cone.
An observer at constant $z<w_0$ with 4-velocity
$\xi^a = (1/\sqrt{g_{tt}})\, \partial/\partial t$ would
measure the following
energy density when the wave front passes:
\[
\rho=T_{ij}\, \xi^i \xi^j=-{1 \over 16\pi t \, g_{tt}}\, 
\delta \,(t+z-w_0)  \quad < \quad 0
\]
which is manifestly negative
and thus violates the weak energy condition.
(Our signature is $(+---)$ as is usual 
for inverse scattering applications, so $g_{tt}>0$. Also we are at
$t>0$.)
Therefore, one would naively think that the present
``null fluid'' is merely an artifact of the singular
transformation and that the ``actual''
one real pole solution has no null fluid. 
(See also work by 
Gleiser and D\'{\i}az\cite{Gleiser,Diaz,Diaz2} on the
problem of
matching solutions in different regions and for methods of
removing the singularities on the light-cone and obtaining smooth
extensions.)

There are at least two ways to find a more physical form 
of the one real pole solutions.
The most obvious is to ``sew together'' the coordinate
patches in a different way, i.e.
to find a better transformation which
is non-singular but still removes the
coordinate infinites which the metric has
in canonical coordinates.
The coordinate transformation
is discussed in section \ref{Disc}.

As an alternative, there seems to be a general belief (\cite{Verd},
sec.\ 3.7.1)
that the one real pole solution be sought as the real limit 
of two complex conjugate poles $w_0 \pm i\epsilon$. The
complex conjugate pole solution is known to be smooth everywhere.
Indeed, this limiting procedure has often been implicitly invoked
to support the validity of simple real pole solutions
such as the cosmic broom.
(The reason one does not simply use the
complex pole solution directly is that it
contains disturbances which travel at
$v>c$, so it in itself does not admit a traveling wave interpretation
similar to that of the real pole solutions, where the shock wave
is strictly located to the light cone and thus
travels at $v=c$.) The limiting
procedure is the subject of section \ref{sec:real}.

Our conclusions are that
\begin{itemize}
\item{
The limiting procedure actually yields a different solution,
so it cannot be used to explain away the null fluid, and
}
\item{
There is no improved transformation, so this avenue
out is also closed.
}
\end{itemize}

\section{Setup}
\label{sec:setup}
We use the conventions and notation of Belinskii and Francaviglia
\cite{Franc}.
The inverse scattering technique is applicable for vacuum metrics of
the form
\be
 \rmd s^2 = f(z,t)(\rmd t^2-\rmd z^2)-g_{ab}(z,t) \,
 \rmd x^a \rmd x^b, \qquad a,b=1,2,
 \label{metric}
\ee
which describes a spacetime admitting an
Abelian $G_2$ group of isometries
\cite{Beli2}. Examples of such metrics are given in the introduction.
This form of the metric is in the ``cosmological case''. The
other (stationary axisymmetric) case, which we do not consider,
is related to this one by a complex transformation, but
the stationary axisymmetric solutions are quite different in character.
In the following $g$ is the $2 \times 2$ matrix representation of
$g_{ab}$.

Belinskii and Francaviglia \cite{Franc} found an explicit
expression for $g$ and $f$ of the complex conjugate pole solution.
Note that we have replaced $\alpha=t$ in their expression.
Also note that the superscripts ``(2)'' and ``(0)''
refer to ``two poles'' and ``no poles'' (background),
respectively. A bar represents complex conjugation.
\begin{eqnarray}
 g^{(2)}_{ab} &=& g_{ab}^{(0)}+\frac{1}{\mathcal D }(\bar{\mu}-\mu)
 (t^2-\abs{\mu}^2)\left[\frac{1}{\mu}(t^2-\mu^2)Q_{22}L_a L_b
 \right. \nonumber \\
 & - & \left.
 \frac{1}{\bar{\mu}}(t^2-\mu^2)Q_{11}\bar{L}_a\bar{L}_b-
 (\bar{\mu}-\mu)Q_{22}Q_{11}g_{ab}^{(0)}\right]
 \label{g2_2}
\end{eqnarray}
\be
  f^{(2)}=c_2 \, {(\mu \bar{\mu} )^3 {\mathcal D} \over
  (t^2-\mu
  \bar{\mu})^2(t^2-\mu^2)(t^2-\bar{\mu}^2)(\mu-\bar{\mu})^2
  t^2} \, f^{(0)}
\label{f}
\ee
where the vector $L_a$ and the scalars $Q_{kl}$ are
\bean
 L_1  &=&  m_1 g_{11}^{(0)} \quad , \qquad
 L_2  =   m_2 g_{22}^{(0)} \quad , \\
 Q_{11} & = &  m_1^2 g_{11}^{(0)}+m_2^2 g_{22}^{(0)}\quad , \\
 Q_{12} & = &  m_1\bar{m}_1 g_{11}^{(0)}+m_2\bar{m}_2 g_{22}^{(0)} = 
 Q_{21}\quad , \\
 Q_{22} & = &  \bar{m}_1^2 g_{11}^{(0)}+\bar{m}_2^2 g_{22}^{(0)} = 
 \overline{Q}_{11}\quad ,
\eean
the $m_1$ and $m_2$ are determined by the seed metric
($s$ is the Kasner parameter):
\begin{equation}
 m_a=
 \left(\ba{cc}
 m_{01}(2w\mu)^{-s}, \quad
 m_{02}(2w\mu)^{s-1}
 \ea\right)
\label{ma}
\end{equation}
with $m_{0b}$ arbitrary complex constants,
and $\mathcal{D}$ is found from
\begin{eqnarray}
 \Delta & = & 
 Q_{11}Q_{22}-Q_{12}^2=t^2(m_1\bar{m}_2-\bar{m}_1 m_2)^2 \nonumber \\
 \mathcal{D} & = & t^2 Q_{11}Q_{22}(\mu-\bar{\mu})^2+
 \Delta(t^2-\mu^2)(t^2-\bar{\mu}^2) \quad.
  \label{eq:D}
\end{eqnarray}
The pole ``trajectory''  $\mu(z,t)$, called trajectory because
it depends on $z$ and $t$,
is
\be
  \mu(z,t)=w - z \pm \sqrt{(w-z)^2-t^2}
\label{traj}
\ee
where $w=w_0 + i\epsilon$. 
(In the original inverse
scattering method, before it was applied to general relativity,
the poles were constants.)
It is of interest to note
that the pole trajectory is the result
of setting the coefficients of any second order poles
to zero, so the method is by construction only treating simple poles.

For the Kasner metric, we note finally
\bean
 g^{(0)}_{11}&=& t^{2s} \quad, \quad g^{(0)}_{22}=t^{2-2s} \ \\
 f^{(0)} &=& t^{2s^2-2s} \qquad.
\eean
Now that the complete metric is given, it
is of interest to count the free parameters:
the complex pole $w$ (eq. \ref{traj}),
the two complex parameters $m_{0a}$ (eq. \ref{ma}) and the
real scaling factor $c_2$ (eq. \ref{f}), making for seven real parameters.

We will also need the expression for the one real pole solution 
\cite{Brad1} (recall the
superscript ``(1)'' indicates one pole):
\bea
g_{ab}^{(1)}&=&\frac{\abs{\tilde{\mu}}}{t}\left(
  g_{ab}^{(0)}+\frac{t^2-\tilde{\mu}^2}{\tilde{\mu}^2 Q}
\tilde{L}_a\tilde{L}_b
\right) \label{eq:onereal} \\
f^{(1)}&=&c_1\frac{\tilde{\mu}^2 Q}{(t^2-\tilde{\mu}^2)\sqrt{t}}
\, f^{(0)}
\nonumber
\eea
with the auxiliary quantities
\begin{eqnarray*}
  \tilde{m}_a &=&(\tilde{m}_{01}\abs{2w_0 \tilde{\mu}}^{-s}, \quad
\tilde{m}_{02}
  \abs{2w_0 \tilde{\mu}}^{s-1}) \\
 \tilde{L}_a &=&(\tilde{m}_1  \, g_{11}^{(0)}, \quad \tilde{m}_2 
\,
g_{22}^{(0)}) \\
 Q&=& \tilde{m}_1^2 \, g_{11}^{(0)}+\tilde{m}_2^2  \, g_{22}^{(0)} \\
 \tilde{\mu}&=&w_0-z \pm \sqrt{(w_0-z)^2-t^2}
\end{eqnarray*}
We will see that these are real parts of the 
corresponding variables for $g^{(2)}$ above, 
e.g. $L_a \rightarrow \tilde{L}_a$ as $\epsilon 
\rightarrow 0$. Again we count the parameters and find
only $\tilde{m}_{0a}$, $c_1$ and $w_0$, 
four real parameters.

\section{The real limit}
\label{sec:real}
\subsection{General conclusions}
The procedure for taking the limit is not unique. For instance, let us
take the expression for the $2 \times 2$ scattering matrix
which generates the transformation,
given by Belinskii and Zakharov \cite{Beli2} as
\be
\chi(\lambda,z,t)=
\openone+\frac{R_1}{\lambda-\mu}+\frac{R_2}{\lambda-\bar{\mu}}
\label{chi}
\ee
where $R_1$ and $R_2$ are $2 \times 2$ matrices and $\lambda$
is the spectral parameter. Now, if
$\mu \rightarrow \bar{\mu}$ as $\epsilon \rightarrow 0$, we would
obtain
\[
\chi(\lambda,z,t)=\openone+\frac{R_1+R_2}{\lambda-\mu} \qquad,
\]  
indistinguishable from the expression for one real pole.
This seems to verify that the limiting procedure
will, indeed, produce a simple real pole.
However, one should note
that the original expression (\ref{chi}) was intended
for distinct poles \cite{Beli2}.

\begin{figure}[h]
   \begin{center}
     \resizebox{2cm}{!}{\includegraphics{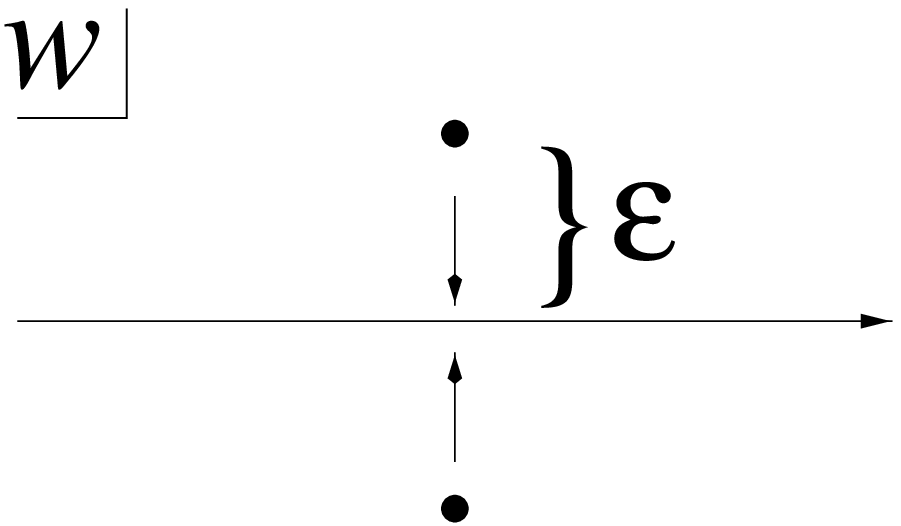}}
     \caption{Taking the limit 
        $\epsilon \rightarrow 0$ in $w=w_0\pm i\epsilon$.}
   \label{fig:limit}
   \end{center}
\end{figure}

Instead of taking the limit in (\ref{chi})
we consider the behavior of the actual metric tensor as
the complex conjugate poles $w_0 \pm i \epsilon$ approach
the real pole $w_0$ by sending $\epsilon
\rightarrow 0$ (see fig. \ref{fig:limit}).

It is useful to discuss
regions I (inside the light cone $(w_0-z)^2<t^2$)
and II (outside the light cone $(w_0-z)^2>t^2$)
separately. Refer to fig. \ref{fig:cone} for the following discussion.

\begin{enumerate}
\item[{\it i)}] {{\bf Region I} \\
{}From equation (\ref{traj}), region I
will never have a real pole trajectory, but instead $g$ approaches the
background. Indeed, from that equation one finds
\[
\abs{\mu}^2 \rightarrow t^2 \qquad \rm{when} \qquad \epsilon
\rightarrow 0
\qquad \mbox{(region I)} \qquad,
\nonumber
\]
which implies that $g$ approaches the background. This
in itself would suggest that we match the solution to the background
in region I.
On the other hand,
\[
g_{tt}=-g_{zz}=f(z,t;\epsilon) 
\rightarrow
\infty \qquad \mbox{as} \qquad {1/ \epsilon^2}
\qquad \mbox{(region I)}
\qquad,
\]
but as long as $\epsilon$ is nonzero, one
can still rescale $f$ using the scaling constant $c_2$
in equation (\ref{f}) to allow for matching.
}
\item[{\it ii)}] {{\bf Region II} \\
We now expand the pole trajectory $\mu$ 
(eq. \ref{traj}) in region II and find
\[
\mu(z,t) \rightarrow \tilde{\mu}(z,t)+i\kappa(z,t)\, \epsilon \qquad
\mbox{(region II)}
\qquad,
\]
where $\kappa(z,t)$ does not approach
zero as $\epsilon \rightarrow 0$.
Thus the complex conjugate pole trajectory
does approach the corresponding
trajectory for the real pole solution.

Now for the metric itself in region II.
Special attention must be paid to the parameters;  
as was noted earlier, simple real pole solutions have only
four real parameters while the complex pole
solution has seven. We know the
pole $w$ becomes real. The two complex parameters $m_{0a}$
must become real in the limit, which accounts for the
difference.
An obvious ansatz for $m_{0a}$ is
\[
m_{0a}=\tilde{m}_{0a} + i\epsilon^p
\]
i.e. to let the complex
parameters approach the corresponding real parameters
as some power $p$ of $\epsilon$.
With this ansatz, it can be seen from equation (\ref{eq:D}) that
\[
{\mathcal D} \rightarrow 0 \qquad
 \mbox{as} \qquad  a\epsilon^2+b\epsilon^{2p}
\qquad \mbox{(region II)} \qquad ,
\]
where $a$ and $b$ do not approach zero.

We find two different cases: for $p<1$, the second term in
(\ref{g2_2}) vanishes, since $\mathcal{D}$ approaches zero slowly
-- we are left with only the background in region II (as well
as region I).
On the other hand, for $p>1$ we do obtain a new solution in
region II.
}
\end{enumerate}

These results differ somewhat from earlier assertions \cite{Franc},
which
state that the parameters must be {\it proportional}
to each other,
otherwise a finite perturbation cannot be found in the limit.
This is again related to the issue of the number of free parameters,
which is reduced when the parameters are taken to be proportional.
Complex conjugate parameters
can never be proportional, so if one takes this claim
literally the limiting procedure can only
be applied to the case of two {\it real} poles. However, in
this argument the
behavior of the parameters $m_{0a}$ was not considered, and
our scheme relies on the
fact that these parameters approach reality with the poles, so it is
free of this restriction. In \cite{GGP} it is shown that a double pole
can be obtained from two coalescing poles if once again the parameters 
are proportional, but the paper does not study the question whether single 
poles can be formed or not.
Neither \cite{Franc} nor \cite{GGP} treated
the possibility of taking the real limit of complex constants.  

Now we turn to discuss the limit in detail.
The general non-diagonal case becomes
rather cumbersome, so we perform the limit numerically.
As an instructive example, we begin by studying
the diagonal case analytically.

\subsection{The diagonal case}
A diagonal perturbation is
obtained by taking, for example, $m_{02}=0$. For a pair of distinct 
complex conjugate poles, we find
\bea
g_{11} &=& {t^2 \over \mu  \, \bar{\mu}} \, g_{11}^{(0)} \, , \qquad
g_{22}={\mu  \, \bar{\mu} \over t^2 } \, g_{22}^{(0)} \, , 
\nonumber
\\
f & = & c_2  \, \frac{\abs{\mu}^6 (g_{11}^{(0)})^2}
{\abs{2w\mu}^{4s}(t^2-\abs{\mu}^2)^2
(t^2-\mu^2)(t^2-\bar{\mu}^2)} \, f^{(0)}
\label{einstein}
\eea
which is known as the Einstein-Rosen metric\cite{Verd}.
At this point, one notes that
the exact same expression applies to the case of two real poles if
$\mu_1$ and $\mu_2$ are substituted
for $\mu$ and $\bar{\mu}$, respectively.
If we take the diagonal perturbation for one real pole, on the other hand,
we find from equation (\ref{eq:onereal}):
\bean
  g_{11} &=& \frac{t}{\abs{\mu}} \, g_{11}^{(0)} \, , \qquad
  g_{22}=\frac{\abs{\mu}}{t} \, g_{22}^{(0)} \, ,  \nonumber \\
  f     &=& c_1  \, \frac{ 
    \mu^2 g_{11}^{(0)}}{\abs{2 w_0 \mu}^{2s}(t^2-\mu^2)\sqrt{t}} \,
f^{(0)}
\eean
By inspection, we cannot obtain this metric from (\ref{einstein}) by
$\epsilon \rightarrow 0$. On the other hand,
the two real pole metric has exactly the form (\ref{einstein}).
Therefore, if the limit is taken at this stage, one
does not obtain a simple real pole but
two confluent real poles.
This conclusion agrees with an older study 
\cite{Verd2}, which was limited to the present
special case of a diagonal perturbation.
The solution is, in general, not the two real pole solution itself
since there is again ambiguity in the way the parameters
coalesce. In addition, the expression for two real poles
was intended for distinct poles, but
this solution clearly exists as a limit.
This is what we mean by ``a pole on a pole''.

\subsection{The non-diagonal case}
\begin{figure}[h]
   \begin{center}
     \resizebox{10cm}{!}{\includegraphics{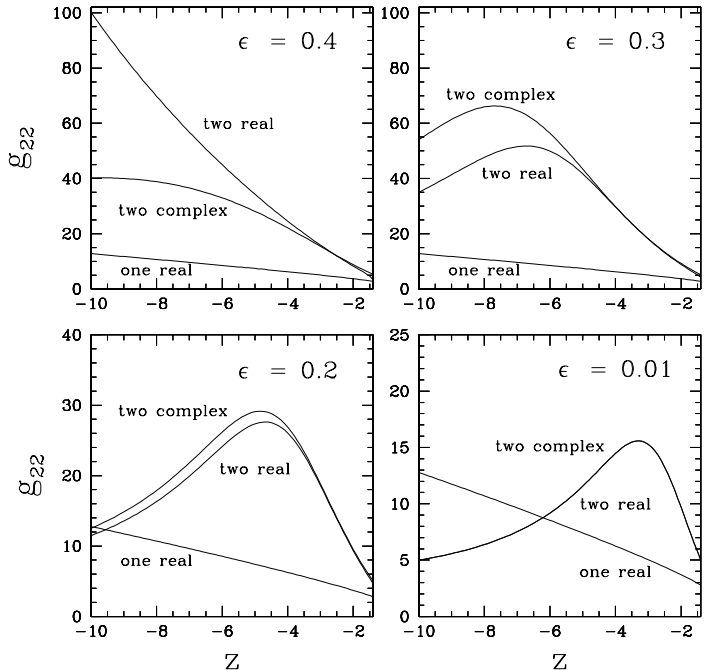}}
     \vspace{-0.7cm}
     \caption{Taking $\epsilon \rightarrow 0$ for $w=1+i\epsilon$,
       $m_{0\mu}=(1+i\epsilon^2, 2+i \epsilon^2)$, $s=0.6$, $t=2$. For the
       two real pole solution we take $w_{\pm}=1\pm \epsilon$,
       $m_{0\mu}^{(\pm)}=(1\pm \epsilon^2, 
       2 \pm \epsilon^2)$ which is a natural choice compared to
       the former.}
       \label{fig:four}
    \end{center}
\end{figure}

We perform the limit numerically in the non-diagonal case.
Details are available elsewhere\cite{Marcus},
but the salient points are summarized here. We find
that conclusions from the diagonal case carry through to
the non-diagonal case.
The metric for complex conjugate poles
approaches the background everywhere for $p<1$, but approaches
a finite perturbation for $p>1$, just
as asserted on analytical grounds earlier by
studying ${\mathcal D}$. An example of this is 
given in fig. \ref{fig:four} for $p=2$. 
The two complex pole solution clearly does not approach
the one real pole solution.

It appears that the two complex pole solution
approaches the two real pole solution in the limit instead.
However, the particular parameter dependence (given in the caption)
has been chosen such that this would happen; in general
the limit is different, although it is true that the limit procedure
shown in fig. \ref{fig:four} may be thought of as the 
most natural limit for this comparison. 
All this is in agreement with the conclusion
in the diagonal case.

It is easy to see numerically that the
metric coefficient $f$ does approach infinity as $\epsilon^{-2}$.
Belinskii and Francaviglia
point out that if the scaling constant $c_2$ is
replaced with (essentially) $\epsilon^2$, we can match $f$ to the
background on the light cone\cite{Franc}. 
This would seem to allow for a finite $f$ in the limit,
since $\lim_{\epsilon\rightarrow 0}\epsilon^2/\epsilon^2=1$.
Unfortunately, when $\epsilon=0$, this
would also mean $c_2=0$. Since there is only one scaling constant,
the metric would then become degenerate in
region II, where $f$ does not approach infinity.

We conclude that the limit of two complex
conjugate poles does produce a solution, 
but it is not the one real pole solution.
Furthermore, this ``pole on a pole'' solution 
also suffers from the deficiency of being infinite
inside the light cone. Thus, the problem of finding a
coordinate transformation which removes the
infinity in $f$ remains for the one real pole solution.

\section{Nonexistence of improved transformations}
\label{Disc}
Instead of the limit procedure above, one could attempt to 
improve the coordinate transformation for a simple real pole
which has been used by several authors \cite{Brad1,FrancCurir}.
A nonsingular transformation, which still does the job
of rendering $f$ finite, would
eliminate the null fluid. 

We here briefly summarize the results from Curir
and one of the authors\cite{Brad1}.
In terms of null coordinates
\[
  u = t + z - w_{0} \, ,  \qquad
  v = t -z + w_{0} \, ,
\]
the metric can be written as
\[
 \rmd s^2 = f \, \rmd u \,  \rmd v -g_{ab} \,
 \rmd x^a \rmd x^b, \qquad a,b=1,2.
\]
When approaching the light cone $u=0$
from the perturbed region II, $f$ goes to infinity as
$1/\sqrt{-u}$ ($v=0$ can be treated in a
similar way). By changing coordinate from $u$ to 
\[
 u'=\mp \frac {c_{1}Q_{0}}{\sqrt{2}}\sqrt{-u} \,
\]
where $Q_{0}$ is the value of $Q$ on the light cone, the metric
changes to
\[
 \rmd s^2 = f' \, \rmd u' \rmd v -g_{ab} \,
 \rmd x^a \rmd x^b, 
\]
where $f'$ is finite on the light cone and matches with $f^{(0)}$
from the unperturbed region I: $f'(u'=0)=f^{(0)}(u=0)$. 
(By defining $u'=u$ in the unperturbed region I, one sees that the 
light cone still is given by $u'=0$ and that $u'$ is continuous
across the light cone.) All metric components are now continuous,
but their first derivatives suffer 
step function jumps across the light cone,
giving rise to delta-functions both in the Weyl tensor and the Ricci
tensor.

So far points with the same value of $u'$ and $v$ have been identified
along the light cone. Could one make the discontinuities disappear by
identifying them in some other way? We have the following result:
\begin{thrm}
There is no (topology-preserving) coordinate transformation of 
the one real pole solution (eq. \ref{eq:onereal})
which both 
makes $f$ everywhere finite
and removes the $\delta$-functions in the Ricci tensor.
\end{thrm}
We proceed to show this. Clearly the light cone is given
by $u'=0$, so the remaining freedom is to identify different $v$ with
each other. But since the 2-dimensional matrix $g_{ab}$ is continuous
across the light cone one finds
\[
g_{abI}(u'=0,v_{I})=g_{abII}(u'=0,v_{II}) \, ,
\]
implying that $v_I=v_{II}$.
New coordinates $\tilde u$ and $\tilde v$ (we choose to perform a
transformation in region II only, since this is general enough) must
then satisfy
\[
\tilde u(0,v)=u'=0 \, ,  \qquad
\tilde v(0,v)=v \, ,
\]
on the light cone. Allowed transformations can then be written as
\[
u'=\tilde u + F(\tilde u, \tilde v) \, , \qquad
v=\tilde v + G(\tilde u, \tilde v) \, ,
\]
where $F$ and $G$ are zero on the light cone
but otherwise arbitrary functions. The requirement that the
metric remains continuous then gives that $F$ and $G$ must approach
zero faster than $\tilde u$. On the other hand, to avoid the
discontinuities giving rise to the delta-function in the Ricci-tensor,
$G$ must approach zero as $\tilde u$ or $F$ must go as 
$\sqrt{-\tilde u}$.
Hence the we cannot avoid
a distributional-valued Ricci tensor 
by this transformation.

We do not consider transformations which induce topology change,
such as rotating the light cone before identifying points. 
Such transformations would create completely different spacetimes
than the ones we consider here.

Our conclusion is that the null fluid is a property of
the one real pole solution which cannot be removed.

\section{Conclusion}
The original idea of eliminating the null fluid in the one real pole
solution through a limit of two complex conjugate poles $w=w_0 \pm
i\epsilon$ is actually impossible to realize.
We reach a different solution, ``a pole on a pole''
instead of a simple real pole solution. Also, we found that
care must be exercised in the taking of the limit since the parameters
must turn real. This is evidenced by the fact that
in one case (what we call $p<1$) the inverse scattering transformation
reduces to the identity transformation.

Additionally, we show that there is no
transformation which accomplishes
the goal of removing the metric infinities
while keeping the null fluid away.
Therefore, barring new topologies,
the simple real pole solution really does
contain a null fluid with negative energy density.
We find this property unphysical in a 
completely classical solution, one
would therefore reject these solutions.
However, a more liberal interpretation
would be that negative energy
on the wave fronts is a characteristic of
cosmological waves, and that this is a sign
that quantum effects in the early universe are required
to produce cosmological shock waves if such waves
are produced at all.

Since the metric in region I approaches the background
under the limiting procedure, we found it natural to match
region II to the background.
It makes the wave interpretation more meaningful since the
metric will be asymptotically Kasner on both sides of the wave front.
It is also common to do so\cite{Verd}.
However, it is of course possible to match
the solution in region II to some metric other than the background.
See e.g. \cite{Griffiths,Nagy} for
treatments where solutions
generated by the inverse scattering technique from Bianchi models
are matched to metrics other than the background. 

Higher-order (i.e. non-simple) poles in the scattering matrix
could also generate viable solutions, but that would be a
different story altogether.

\section{Acknowledgements}

We wish to thank Lawrence C. Shepley for valuable structural advice,
Matthew W. Choptuik for constructive criticism, and Lennart Stenflo
at Ume\aa\,University for hospitality.

\end{document}